\documentstyle[12pt]{article}
\begin{document}
\begin{center}
\textbf{THE  CASIMIR  PROBLEM  OF  SPHERICAL  DIELECTRICS:  A  SOLUTION  IN  
TERMS  OF  QUANTUM  STATISTICAL  MECHANICS}\footnote{Written for a festschrift
issue of {\it Journal of Statistical Physics }, dedicated to George Stell.}\\
\bigskip
\bigskip
J. S. H{\o}ye$^2$ and I. Brevik$^3$ \\
\bigskip
$^2$Department of Physics, Norwegian University of Science and Technology,\\
N-7491 Trondheim, Norway\\
\bigskip
$^3$Division of Applied Mechanics, Norwegian University of Science and
Technology,\\
N-7491 Trondheim, Norway
\bigskip
March 1999\\
\end{center}
\bigskip
\bigskip
\begin{abstract}
The Casimir energy for a compact dielectric sphere is considered in a novel
way, using the quantum statistical method introduced by H{\o}ye-Stell and
others.  Dilute media are assumed. It turns out that this method is a very
powerful one: we are actually able to derive an expression for the Casimir
energy that contains also the negative part resulting from the attractive van
der Waals forces between the molecules. It is precisely this part of the
Casimir energy that has turned out to be so difficult to extract from the
formalism when using the conventional field theoretical methods for a
continuous medium. Assuming a frequency cutoff, our results are in agreement
with those recently obtained by G. Barton [J. Phys. A: Math. Gen. {\bf 32}, 525
(1999)].
\end{abstract}


\newpage
\section{Introduction}
It is a pleasure to contribute this article to a festschrift volume for
Professor Stell. The article is one in a series of articles published by the
present authors on the Casimir effect and related topics, using methods of
statistical mechanics for quantized systems at thermal equilibrium. Besides
contributions from others, these methods were developed by one of the authors
(JSH) in cooperation with Professor Stell in their extensive studies of polar
and polarizable fluids through several years. This long-lasting cooperation,
which was initiated in 1972, is still active today.

The Casimir energy problem for a compact spherical ball is a many-facetted
problem; the formal solution of it is to an unusual degree dependent on the
mathematical method of approach chosen. The Casimir effect as such is now a
well-known effect in physics \cite{casimir48}. It is ordinarily examined with
the use of field theory in dielectric media, allowing the medium to possess a
refractive index $n$ (even dispersive effects can in principle be dealt with in
this way, if $n$ is assumed to depend on the frequency). The standard
configuration does not involve curved boundaries at all, but consists instead
of two plane plates separated by a small gap. In this geometrical configuration
the phenomenological electromagnetic theory, as constructed mainly by Lifshitz
\cite{lifshitz55} is fully adequate, and leads to a prediction for the Casimir
force between the plates that has recently been verified experimentally to an
impressive accuracy of about one per cent \cite{lamoreux97}, \cite{mohideen98}.

If we now leave the parallel-plate configuration and consider instead a single
dielectric ball, the situation becomes much less clear-cut. The history along
this direction of research may be taken to start with the calculation of Boyer
on a singular perfectly conducting {\it shell} \cite{boyer68}: he found the
Casimir energy $E$ to be positive, corresponding to an {\it outward} directed
surface force. Later, the dielectric ball was considered by Milton
\cite{milton80}, \cite{milton96},\cite{milton99}, Milton and Ng
\cite{milton97},\cite{milton98},  Brevik {\it et al.}
\cite{brevik94},\cite{brevik98},\cite{brevik98a},\cite{brevik99} and several
others. Some consensus seems by now to have been reached as regards the Casimir
energy $E$ as found by field theoretical methods: this energy is positive,
corresponding to a repulsive force, and is given by
\begin{equation}
E=\frac{23}{384}\frac{\hbar c}{\pi a}(n-1)^2,
\end{equation}
\label{1}
for a dilute sphere whose radius is $a$.

Faced with this field theoretical result one becomes however surprised, for the
following physical reason: the Casimir energy should be the cooperative result
of the van der Waals forces between the molecules in the ball. The van der
Waals forces are necessarily {\it attractive}. How can these forces sum up to
give a {\it repulsive} total surface force? The natural answer to this question
is that the field theoretical calculation, based as it is on a continuum model
for the dielectric, is unable to copy with the attractive part. In other words,
the attractive terms are necessarily lost in the regularization process. An
important progress was recently made by Barton \cite{barton99}; he made use of
quantum mechanical perturbation theory to  second order, imposed an exponential
cutoff in wave numbers, and arrived at a definite expression for the Casimir
energy containing also the cutoff dependent, attractive (and actually also
repulsive) terms. Moreover, a cutoff independent, repulsive term was contained
in the energy expression, which was in precise agreement with Eq.(1) above.
There are actually some indications of the same kind already in the paper of
Milton and Ng \cite{milton98}: they derived the cutoff independent Casimir
energy starting from the van der Waals forces, omitting the divergent terms.

And this brings us to the central theme of the present paper, namely to
rederive the expression for the Casimir energy using the perhaps somewhat more
unconventional quantum statistical methods that were developed by H{\o}ye and
Stell, and others. Central references for the present work are \cite{hoye81}
and \cite{hoye82}. Others that also include evaluation of frequency spectra are
\cite{chandler81}. As we will see, this method is very powerful, and we will be
able to establish contact with the results of Barton. The line of development
of the application of this method to the Casimir problem is the following: 
Some years ago Brevik and H{\o}ye \cite{brevik88} showed that the Casimir
energy between two point particles is the same as the free energy due to two
quantized fluctuating dipole moments interacting via the dipolar radiation
interaction (zero frequency limit corresponds to the static dipole - dipole
interaction). Later, H{\o}ye and Brevik \cite{hoye98} extended this method to
evaluate the Casimir force between a pair of parallel dielectric plates
separated by a small gap. Performing this more complex calculation with the use
of statistical mechanics for systems in thermal equilibrium, we were able to
rederive the known results.

Below we will evaluate the free energy in a dielectric dilute medium, again
using the same methods of statistical mechanics. Based on this we will make
contact with the results of Barton, as mentioned, as well as with the results
obtained in field theory. On the basis of our method the physical origin of the
divergences is easily understood. The problem, as anticipated above, has its
origin in a continuum description of dielectric media, while a realistic system
has to have a microscopic structure involving a minimum separation between
molecules due to repulsive cores. 

\section{Basic formalism}
We begin by recapitulating some of the basic formulas from our earlier work
\cite{brevik88}. For a pair of polarizable particles the free energy $F$ due to
their mutual interaction is 
\begin{equation}
-\beta F = \frac{3}{2}\sum_K \alpha_K^2(2 \psi_{DK}^2(r) + \psi_{\Delta
K}^2(r));
\end{equation}
\label{2}
cf. Eq.(5.14) in \cite{brevik88}. Here $\alpha_K$ is the frequency dependent
polarizability, and
\begin{equation}
K=2\pi n/\beta
\end{equation}
\label{3}
is the Matsubara frequency related to the frequency $\omega$ via
\begin{equation}
K=-i\hbar \omega.
\end{equation}
\label{4}
Further, $n$ is an integer, $\beta=1/k_B T$ is the inverse temperature, and
$\psi_{DK}$ and $\psi_{\Delta K}$ are the two radial parts of the radiating
dipole - dipole interaction as given by Eqs.(5.9) and (5.10) in
\cite{brevik88}. Performing the sum in (2) for $\beta \rightarrow \infty$ (i.e.
$ T \rightarrow 0$), we obtain Eq.(5.16) in \cite{brevik88} 
\begin{equation}
F=-\frac{23 \hbar c \alpha^2}{4\pi r^7},
\end{equation}
\label{5}
which is the known result for the Casimir effect.

For a low density medium the total free energy $\Delta$ can now be obtained by
summing or integrating (2) over pairs of particles in a volume $V$ such that
\begin{equation}
\Delta=\frac{1}{2}\rho^2 \int d{\bf r}_1 d{\bf r}_2\, F,
\end{equation}
\label{6}
where $\rho$ is the number density and ${\bf r}={\bf r}_1-{\bf r}_2$. At $T=0$,
(5) is to be inserted. Clearly, the integral will diverge, due to the behaviour
for small values of $r$. However by integrating over a small sphere a finite
term, which is positive, can be separated out; cf. \cite{brevik98a}. This
finite term turns out to coincide with the field theoretical result. We will
show below that the divergences found using other kinds of approach are
connected with this small $r$ behaviour. Equation (2), together with (6), will
be shown to lead to results in agreement with those obtained from quantum
mechanical perturbation theory to second order \cite{barton99}. The exponential
cutoff used by Barton in Fourier space can be introduced also in our approach;
it corresponds to a small "soft" $r$ cut out from the otherwise continuous
medium, and will be a rough approximation to real systems. As mentioned above,
real systems are not continuous but consist of molecules that have a minimum
separation due to hard cores.

\section{Calculation of the free energy}
Let us now calculate the free energy $\Delta$, as given by (6). In order to
establish connection with  perturbation theory, we first represent (2) in terms
of Fourier quantities for which, as we will see, a wave vector cutoff can be
introduced easily. The radiating dipole - dipole interaction used in (2) can be
written as 
\begin{equation}
\psi(12)=\psi_{DK}(r)D_K(12)+\psi_{\Delta K}(r)\Delta_K(12),
\end{equation}
\label{7}
with 
\begin{eqnarray}
D_K(12)           &=&
3(\hat{r}\,\hat{a}_{1K})(\hat{r}\,\hat{a}_{2K})-\hat{a}_{1K}\,\hat{a}_{2K},
\nonumber  \\    \Delta_K(12)      &=& \hat{a}_{1K}\hat{a}_{2K}. \nonumber
\end{eqnarray}
Here the hats denote unit vectors, and ${\bf a}_{iK}$ is the Fourier transform
of the fluctuating dipole moment of particle number {\it i}  in imaginary time;
cf. Eq.(5.2) in \cite{brevik88}. Equation (7) can be Fourier transformed to
give
\begin{equation}
\tilde{\psi}(12)=\tilde{\psi}_{\Delta K}(k)\tilde{D}_K(12)+\tilde{\psi}_{\Delta
K}(k)\Delta _K(12),
\end{equation}
\label{8}
with
\begin{eqnarray}
\tilde{D}_K(12)   &=& 3(\hat{k}\,\hat{a}_{1K})(\hat{k}\,
\hat{a}_{2K})-\hat{a}_{1K}\,\hat{a}_{2K}, \nonumber \\
\psi(12)          &=& \frac{1}{(2\pi)^3}\int {\tilde{\psi}}(12)e^{i{\bf k}{\bf
r}} d {\bf k}.  \nonumber
\end{eqnarray}
With this Eq.(2) can be rewritten as
\begin{equation}
-\beta F=\frac{3}{2}\frac{1}{(2\pi)^6}\sum_K\int M_K e^{i({\bf (k+k') r}}d{\bf
k}d{\bf k'}
\end{equation}
\label{9}
where
\begin{equation}
M_K=\tilde{\psi}_{\Delta K}(k)\tilde{\psi}_{\Delta
K}(k')(3(\hat{k}\hat{k'})-1)+
\tilde{\psi}_{\Delta K}(k)\tilde{\psi}_{\Delta K}(k').
\end{equation}
\label{10}
Like expression (2), this is obtained after orientational averaging of the
products of terms containing $\tilde{D}_K(12)$ and $\Delta_K(12)$ with respect
to $\hat{a}_{iK}$.

To get further the explicit Fourier transformed interaction terms are needed.
These follow from the solution of Maxwell's equations, and like the
corresponding terms in (2) (Eq.(5.10) in \cite{brevik88}) they were used by H\o
ye and Stell when dealing with the refractive index of fluids \cite{hoye82}.
Thus from Eq.(7) in \cite{hoye82} we have \footnote{Note here that the Fourier
transform in imaginary time on the interval from $0$ to $\beta$ is the same
function as the real time transform; cf. the derivation in Appendix B of
\cite{brevik88}.}
\begin{equation}
\tilde{\psi}_{DK}(k)=-\frac{4\pi}{3}\frac{k^2}{k^2-\omega^2},~~~~
\tilde{\psi}_{\Delta K}(k)=\frac{4\pi}{3}\left( \frac{2k^2}{k^2-\omega^2}-
(2+\Theta) \right) ,  
\end{equation}
\label{11}
with $K =-i\hbar c\omega$. (For simplicity $\omega$ is replaced by $c\omega$ 
where $c$ is the light velocity.) Here the parameter $\Theta$ introduced by H\o
ye and Stell \cite{hoye76} is used.  A purpose to introduce it was by
$\gamma$-parametrization of the dipole-dipole interaction to obtain a
continuous family of mean field theories ($\gamma \rightarrow 0$) of polar
fluids. Here $\gamma$ is the inverse range of $\psi_{\Delta K}(r)$, and for
$\psi_{DK}(r)$ it is the inverse range inside which the dipolar $1/r^3$
behaviour is cut or rounded off. As seen from (11) the $\Theta$ is thus the
integrated amplitude of the $\psi_{\Delta K}$-term ($\omega =0$). This
parameter was also used in \cite{hoye82}, part IV, and its Eq.(57) for the
direct correlation function corresponds to Eq.(11) here. With its Eqs.(56) and
(59) the dielectric constant $\varepsilon$ can then in general be expressed as
(for small $\omega/\gamma \rightarrow 0$)
\begin{equation}
\frac{\varepsilon -1}{(1-\Theta)\varepsilon+(2+\Theta)}=\frac{4\pi}{3}\rho
\alpha,
\end{equation}
\label{12}
where $\rho$ is the number density of particles. (Here a possible density
dependence of $\alpha \rightarrow \alpha_{eff}$ which is proportional to the
fluctuating dipole moment squared, is disregarded.)   

The separate term $(2+\Theta)$ at the end of (11) will necessarily yield
infinity when inserted in (9) and summed. In ${\bf r}$-space it gives a
$\delta$-function at ${\bf r}=0$. As pairs of particles in reality are
separated, we can simply remove  this term here. This amounts to putting
$2+\Theta=0$, by which 
\begin{equation}
\gamma = \frac{\varepsilon -1}{\varepsilon}=4\pi \rho \alpha.
\end{equation}
\label{13}
Note here that this choice is consistent with the continuum approach ($ \gamma 
\rightarrow  \infty  $), where only a transverse radiating field is implicitly
considered. That is, with $\Theta = -2$ the longitudinal part vanishes, as
follows from Eq.(28) in \cite{hoye82}. Then one has (with the direct
correlation function $c \rightarrow \psi$)
$\tilde{\psi}_1=\frac{1}{3}(\tilde{\psi}_{\Delta K}+2\tilde{\psi}_{DK})=0$,
while the transverse part becomes
$\tilde{\psi}_2=\frac{1}{3}(\tilde{\psi}_{\Delta
K}-\tilde{\psi}_{DK})=(4\pi/3)k^2/(k^2-\omega^2)$.

Inserting (11) into (10) we obtain
\begin{eqnarray}
M_K  &=& \frac{(4\pi)^2}{3}\frac{k^2}{k^2-\omega^2}\frac{k'^2}{k'^2-\omega^2}[
(\hat{k}\,\hat{k}')^2+1] \nonumber \\
     &=& \frac{(4\pi)^2}{3}\frac{k^2\,k'^2}{k^2-k'^2} \left[
\frac{1}{k'^2-\omega^2}-\frac{1}{k^2-\omega^2} \right]
         \left[ (\hat{k}\,\hat{k}')^2+1 \right].
\end{eqnarray}
\label{14}
Now the summation in (9) can be considered, and by restricting ourselves to
frequency independent polarizability $\alpha_K=\alpha$ we can easily perform
the sum since (14) is of standard form for simple harmonic oscillators
\cite{hoye81}. We have ($K=2\pi n/\beta$)
\begin{eqnarray}
\sum_K \frac{(\hbar \omega_0)^2}{(\hbar \omega_0)^2+K^2}&=&\frac{1}{2}\beta
\hbar \omega_0
                                  \; \frac{\cosh (\frac{1}{2}\beta \hbar
\omega_0)}{\sinh (\frac{1}{2}
                                    \beta \hbar \omega_0)} \nonumber \\
                                                         & & \stackrel{\beta
\rightarrow \infty}{\longrightarrow}
                                    \frac{1}{2}\beta \hbar \omega_0.
\end{eqnarray}
\label{15}
With $\alpha_K=\alpha$, use of (15), and (14) inserted for $M_K$, we now easily
find at $T=0$ ($K=-i\hbar c \omega,~~\hbar \omega_0\rightarrow \hbar ck $ and
$\hbar ck'$)
\begin{eqnarray}
\sum_KM_K &=& \frac{(4\pi)^2}{3}\frac{1}{k^2-k'^2}\frac{1}{2}\beta \hbar
c(k'k^2-kk'^2)[(\hat{k}\,\hat{k}')^2+1]
                                          \nonumber \\
          &=& \beta \hbar c
\frac{(4\pi)^2}{6}\frac{kk'}{k+k'}[(\hat{k}\,\hat{k}')^2+1].
\end{eqnarray}
\label{16}
When inserting this into (9), and further inserting into (6), we see that the
result will diverge due to the small $r$ (or large $k$) behaviour. This
divergence can be avoided by introducing a large wave number cutoff, as Barton
did \cite{barton99}. Thus we incorporate a factor $\exp(-\lambda k)$ in the
interaction terms (11), which implies a factor $\exp (-\lambda (k+k'))$ in
(16). Regarding the electromagnetic field as a set of harmonic oscillators that
mediate the interaction between the particles, the effect of this cutoff is to
remove the high frequency oscillators. With (16) and (9) inserted into (6) we
then obtain
\begin{equation}
\Delta=-\frac{\gamma^2}{2(16\pi^3)^2}\int d{\bf r}_1 d{\bf r}_2\int d{\bf
k}d{\bf k'} k k'
e^{i({\bf k}+{\bf k'}){\bf r}}\;\frac{e^{-\lambda
(k+k')}}{k+k'}[(\hat{k}\,\hat{k}')^2+1],
\end{equation}
\label{17}
where $\gamma$ is given by (13).

 With $\Delta=\gamma^2\Delta_2$ this is precisely the result obtained by Barton
\cite{barton99} applying quantum mechanical perturbation theory to the
dielectric continuum. That is, we have recovered the second order contribution
which in the case of a scalar field is given by Eq.(4.2) in \cite{barton99}
where the equality
\begin{equation}
\frac{\exp(-\lambda (k+k'))}{k+k'}=\int_{\lambda}^{\infty}d\xi \exp (-\xi
(k+k'))
\end{equation}
\label{18}
has been used. When dealing with the electromagnetic field, as we do here, the
result in \cite{barton99} is modified with the factor
$[(\hat{k}\,\hat{k}')^2+1]$ as given by Eq.(B.2) in \cite{barton99}. With this,
it is seen that our results are identical with those of Barton. Thus, further
considerations based upon this basic agreement will necessarily be the same,
and will not be repeated here.

\section{Further remarks}
 
Let us  make a few remarks on the resulting free energy (or internal energy at
$T=0$) for a spherical body of radius $a$ where a positive cutoff independent
term going like $1/a$ shows up also within the field theoretical approach. This
is commonly interpreted as equivalent to a repulsive Casimir surface force.
From our approach it is now obvious that this term reflects the large $r$-
behaviour of the free energy (5) of particle pairs. For a finite system such as
a sphere the resulting free energy will in general be {\it larger} than its
bulk value since there is no material outside with which it can interact. This
missing interaction first of all manifests itself in a positive term that
reflects the surface tension and is proportional to the surface area. This term
is present in the final result given in the Abstract in \cite{barton99}. In
addition, there is the $1/a$ term which can be associated with the $1/r^7$-tail
in connection with the  missing material away from the surface. 

As another point, let us note that in \cite{barton99} there is an additional
leading term due to first order perturbation theory. This term is not present
in our derivation above. However it can be identified in a straightforward
manner from the earlier work of H\o ye and Stell \cite{hoye82}. It represents a
self-energy of the electromagnetic field attached to the polarizable particles.
As such it is just part of the properties of isolated single particles, and
should accordingly not be included in the free energy density above. In fact,
it is part of the radiation reaction of accelerated charges and is incorporated
in the resulting physical momentum. Unfortunately this momentum correction for
a classical particle is infinite , so the "bare" mass of particles is negative
resulting in the well known "runaway" problem in classical electromagnetism
\cite{jackson75}. This also reflects itself in the refractive index problem
considered in \cite{hoye82}. Further, H\o ye and Lomba made numerical
calculations to obtain a minor non-causal tail in the dielectric response of
the fluid ( i. e., a minor response will appear before an electric field is
applied to avoid exponential growth or runaway for increasing time
\cite{hoye90}).

With cutoff in wave vectors the above mass correction can be made finite. The
term of interest is then the component $\tilde {\psi}_{\Delta K}(k)$ of Eq.(8).
Including the shielding factor in (11) we then have (with $2+\Theta=0$ as
above)
\begin{equation}
\tilde{\psi}_{\Delta K}(k)=\frac{4\pi}{3}\frac{2k^2}{k^2-\omega^2}\,e^{-\lambda
k}.
\end{equation}
\label{19}
As shown by Eqs.(12)-(15) of \cite{hoye82}, this gives rise to a
self-interaction (for given $K, ~~s \rightarrow a_K$)
\begin{equation}
\Delta \phi(a_K)=-\frac{1}{2}a_K^2\psi_{\Delta K}(0),
\end{equation}
\label{20}
where the minus sign is due to the definition used. The total internal energy
contribution that follows from this is
\begin{equation}
\Delta_1=\sum_K \rho V \langle \Delta \phi(a_K) \rangle = -\frac{1}{2}\rho V
\sum_K
\langle a_K^2 \rangle \psi_{\Delta K}(0),
\end{equation}
\label{21}
where $\rho V$ is the number of particles in a volume $V$.

Now $\langle a_K^2 \rangle=3\alpha_k /\beta$; cf. \cite{brevik88}. So with
$\alpha_K=\alpha$ we obtain by first using (15) for $\beta \rightarrow \infty$
\[ \sum_K\tilde{\psi}_{\Delta K}(k)=\frac{4\pi}{3}\beta \hbar ck e^{-\lambda
k}, \]
or
\begin{equation}
\sum_K \psi_{\Delta K}(0)
=\frac{4\pi}{3}\frac{\beta \hbar c}
{(2\pi)^3}\int 
k e^{-\lambda k}\,d{\bf k}
=\frac{4}{\pi}\frac{\beta \hbar c}{\lambda^4},
\end{equation}
\label{22}
which together with (13) and (21) yield 
\begin{equation}
\Delta_1=-\frac{1}{2}\rho V \frac{3\alpha}{\beta}\frac{4}{\pi}\frac{\beta \hbar
c}{\lambda^4}=
-\gamma \frac{3}{2\pi^2}\frac{\hbar c}{\lambda^4} V.
\end{equation}
\label{23}
This is precisely the first order contribution obtained by Barton
\cite{barton99}. As mentioned above, this energy is a quantity that is part of
the free particles themselves, and can thus not be separated out. So the free
energy of the interacting system of polarizable particles will not include this
term.

After the above was written, we have become aware that the recent paper of
Bordag, Kirsten, and Vassilevich \cite{bordag99} arrives at results closely
related to those obtained by Barton \cite{barton99}. The authors of
\cite{bordag99} make use of field theoretical path integral methods which are
in themselves quite different from those used in \cite{barton99} as well as in
the present paper, but the expression for the Casimir energy obtained in
Section IV in \cite{bordag99} nevertheless parallels that obtained in
\cite{barton99}. (There are some differences in numerical coefficients of the
divergent terms, due to different regularization methods employed.) We thank
the authors of \cite{bordag99} for making us aware of this correspondence.

\end{document}